\documentclass[prb,twocolumn, showpacs,preprintnumbers,amsmath,amssymb]{revtex4}
\usepackage{graphicx}
\usepackage{epsfig} 
\usepackage{color}
\usepackage{epstopdf}
\usepackage{bm}
\newcommand{\figurewidth}{3.4in}
\newcommand{\FG}[1]{Fig.~\ref{#1}}
\newcommand{\EQ}[1]{Eq.~(\ref{#1})}

\newcommand{\rtan}{|\bm r - \bm \tau_{\alpha n}|}
\newcommand{\op}[1]{\widehat{\rm #1}}
\newcommand{\pu}{\hspace{0.4cm} .}
\newcommand{\ko}{\hspace{0.4cm} ,}
\newcommand{\bra}[1]{{\langle\, #1 \,\vert\,}}
\newcommand{\ket}[1]{{\,\vert\, #1 \,\rangle}}
\newcommand{\e}[1]{{\rm e}^{#1}}
\newcommand{\difr}{\;{\rm d}^3 {\bm r}}
\newcommand{\basis}{{\bm \tau}_{\alpha n}}
\newcommand{\argg}{({\bm G})}

\begin{document}
\title{Atomic Effective Pseudopotentials for  Semiconductors}

\date{\today}
\author{J. R. C\'ardenas}
\author{G. Bester}
\email[Email: ]{g.bester@fkf.mpg.de}
\affiliation{Max-Planck-Institut f\"ur Festk\"orperforschung, Heisenbergstr. 1, D-70569 Stuttgart, Germany}

\begin{abstract}
We derive an analytic connection between the screened self-consistent effective potential from density functional theory (DFT) and atomic effective pseudopotentials (AEPs). The motivation to derive AEPs is to address structures with thousands to hundred thousand atoms, as given in most nanostructures. The use of AEPs allows to bypass a self-consistent procedure and to address eigenstates around a certain region of the spectrum (e.g., around the band gap). The bulk AEP construction requires two simple DFT calculations of slightly deformed elongated cells. The ensuing AEPs are given on a fine reciprocal space grid, including the small reciprocal vector components, are free of parameters, and involve no fitting procedure. We further show how to connect the AEPs of different bulk materials, which is necessary to  obtain accurate band offsets. We derive a total of 20 AEPs for III-V, II-VI and group IV semiconductors and demonstrate their accuracy and transferability by comparison to DFT calculations of strained bulk structures, quantum wells with varying thickness, and semiconductor alloys. 
\end{abstract}



\maketitle
\section{Introduction}

The main motivation for deriving atomic effective pseudopotentials (AEPs) relies on the computational savings that may be gained by circumventing the self-consistent optimization of the density, which is at the heart of density functional theory (DFT). The aim is thereby to address the omnipresent nanoscale, where the relevant structures include thousands, and often hundred thousands, of atoms. The philosophy behind an AEP, not only bypasses the need for a self-consistent solution, but also allows to focus on a selected part of the eigenvalue spectrum. The number of ``bands" that scales with the number of atoms in DFT, is now independent of the system size. This latter characteristic is ideal for the study of optical properties or transport, where mainly the energy states around the band gap of the materials are involved.

The idea to replace the strong Coulomb potential of the nucleus together with the bound electrons by an effective, weaker (pseudo)potential is not new  \cite{fermi34,hellmann35}. The first generation of the method\cite{phillips58,brust62,brust62b,cohen66,chelikowsky76,cohen97} was based on the fitting of an empirical pseudopotential constructed to reproduce experimentally determined energy levels of bulk crystals. In this procedure, known as the empirical pseudopotential method (EPM), the entire band structure as well as optical properties of bulk solids can be obtained with good accuracy with only very few Fourier components of the pseudopotential (in the order of three). These few parameters were simply used as fitting parameters and adjusted to reproduce known quantities. 

In the original EPM, the parameters were directly linked to the investigated bulk structure through the use of certain $\bm G$-vectors and transferability (i.e. the applicability of one potential to different lattice constants or different structures) was not supported by this type of construction. A simple extension via interpolation to neighboring points in $\bm G$-space \cite{caruthers78,andreoni78} or a full analytic dependence of the pseudopotential on the length of the $\bm G$-vectors \cite{xia88,wang94b,mader94} was a natural extension of the method. With these continuous empirical pseudopotentials, together with massively parallel codes \cite{wang94,wang96,wang99a} or with favorable basis sets \cite{wang99b,esseni05,bester09},  the electronic and optical properties of nanostructures could be accurately calculated \cite{williamson00,franceschetti99,bester03,bester04,califano04b,ediger07a,bester09}, as well as transport properties \cite{Esseni11,jiang11}. 

With the aim of getting an improvement in the quality of the wave functions, and hence of the transferability, a second generation of empirical pseudopotentials was developed by Wang, Fu and Zunger\cite{wang95,fu97b}. In this approach, a series of bulk self-consistent DFT calculations, in the local density approximation (LDA), are performed using a few different structures (zincblende, wurtzite, ...) at different lattice constants. The screened local effective potentials from DFT are transformed to $\bm{G}$ space and fitted by a set of Gaussian functions. This results in a continuous representation of the atomic empirical potentials in $\bm{G}$ space with a good transferability between different bulk structures. However, as the structures used in the potential generation are all bulk systems, no information on the long range (small $\bm G$-vector) inter-atomic interaction is available. Hence the potentials at $\bm G$-vectors shorter than $2\pi$ divided by the lattice constant, remains unknown. Unfortunately, it is precisely this range of $\bm G$-vectors that becomes relevant for the calculation of nanostructures, where the shortest relevant $\bm G$-vectors is given by $2\pi$ divided by the cluster size. In the original work \cite{wang95,fu97b}, the Gaussian function was interpolated from the shortest known $\bm G$-vector (2$\pi/a$) to the $\bm G$=0 point. The potential at the $\bm G$=0 was adjusted to reproduce experimentally known work functions \cite{wang95} and later using an atomic model potential and screening \cite{fu97b}. Furthermore, the energy cut-off was significantly reduced to save computational resources, so that an empirical Gaussian function had to be introduced to compensate for this reduction. Although the quality of the ensuing semiempirical pseudopotentials (SEPM) must be expected to be significantly higher than the EPM, these difficulties may have been the reason for the rather limited use of this potential generation. 

In this paper, we introduce a new generation of pseudopotentials following the philosophy of the EPM and SEPM to derive a pseudopotential that represent the scattering due to all the electrons. In this sense, it is distinct from the ionic pseudopotentials derived for DFT \cite{hamann79a,vanderbilt85a,bachelet82a,troullier90a,kerker80a,vanderbilt90a,blochl90a,blochl94}, although we use a separable formalism to treat the non-local part of the potential stemming from this community \cite{kleinman82a}.
We derive our atomic effective pseudopotentials (AEPs) via an analytic connection to the effective crystal potential from DFT. The use of large and judiciously chosen slightly deformed supercells gives us access to the long range response of the potential. We therefore derive the pseudopotentials on a very dense grid of $\bm{G}$ vectors, including both, the long and short range interaction regimes. The method is free of parameters and does not involve any fitting procedure and is as {\it ab initio} as the underlying DFT calculations. This simplicity allows us to generate, with a minimum of effort, unique (for a certain DFT norm-conserving pseudopotential) AEPs. 
We have successfully generated AEPs for 20 compound semiconductor materials.
A comparison between the results of our code LATEPP\cite{bester12} using our AEPs and DFT\cite{gonze02} shows differences for the bulk calculations in the order of a few tenth of meV, which originates from our spherical approximation of the local part of the potential. We have tested the transferability of the potential on different types of quantum wells and find very satisfying results, demonstrating a high level of transferability. 

\section{Methodology}

Our methodology is based on DFT\cite{hohenberg64a,kohn65a} where the central quantity is the effective Kohn-Sham potential $V_{\rm eff}$ that describes all the interactions of a single electron with its environment:
\begin{eqnarray}
\label{eq:ks} 
&&\left( -\frac{\hbar^2}{2m}\Delta + V^{\rm eff}(\bm r) \right) \psi_i(\bm r) = \varepsilon_i \psi_i(\bm r) \nonumber\\ 
&& V^{\rm eff}(\bm r)  = V^{\rm ext}(\bm r) + V^{\rm Hartree}[n(\bm r)] + V^{\rm xc}[n(\bm r)]  \label{eq:effpot} \\
&& n(\bm r) = \sum_i^{\rm occ} |\psi_i(\bm r)|^2 \quad , \nonumber
\end{eqnarray}
where $n(\bm r)$ is the charge density of all occupied single-particle states $\psi_i$. The second key component is given by the transformation from the true Coulomb potential 
\begin{equation}
V^{\rm ext}(\bm r) \equiv  V^{\rm ext}(\bm r; \{\bm R_\alpha\}) =  -\sum_{\alpha}^{N_c}\frac{Z_\alpha e^2}{|\bm r - \bm R_\alpha|}
\end{equation}
into a pseudopotential $\widehat{V}^{\rm psp}$. The core electrons are thereby frozen in their atomic configuration following the frozen-core approximation \cite{barth80a}. 
The pseudopotential is usually written as a sum of angular momentum projectors:
\begin{eqnarray}
\widehat{V}^{\rm psp} & = & \sum_{\alpha}^{N_{\rm species}}\sum_n^{N_\alpha} \sum_{lm} v_{\alpha l} (\rtan) \op{P}_{lm} \ko
\end{eqnarray}
where $\alpha$ describes the atom type and runs from one to $N_{\rm species}$ and $n$ describes the atom number running from one to the number of atoms of type $\alpha$, $N_\alpha$. The atomic positions are given by $\basis$ and $\op{P}_{lm}$ is the projection operator. Here we note that all the pseudopotentials describe correctly the long-range Coulomb interaction, beyond a certain cut-off:
\begin{equation}
\label{eq:potdecay}
v_{\alpha l}(r) = -\frac{eZ_\alpha}{r} \quad \mbox{for} \quad r > r_{\alpha l}^{\rm cut} \ko
\end{equation}
where $Z_\alpha$ is the charge of the valence of the respective atom. It is therefor customary to perform the following transformation: 
\begin{widetext}
\begin{eqnarray}
\widehat{V}^{\rm psp} & = & \sum_{\alpha n} \sum_{lm} v^{\rm psp,loc}_{\alpha } (\rtan) \op{P}_{lm} + 
\sum_{\alpha n}\sum_{lm}[v_{\alpha l} (\rtan) -  v^{\rm psp,loc}_{\alpha } (\rtan)] \op{P}_{lm}  \nonumber \\
& = & \sum_{\alpha n} v^{\rm psp,loc}_{\alpha } (\rtan) \sum_{lm}\op{P}_{lm} + 
\sum_{\alpha n}\sum_{lm}[v_{\alpha l} (\rtan) -  v^{\rm psp,loc}_{\alpha } (\rtan)] \op{P}_{lm}\label{eq:vpsp} \\ 
& = & \sum_{\alpha n} v^{\rm psp,loc}_{\alpha} (\rtan) + 
\sum_{\alpha n}\sum_{lm} \delta v_{\alpha l} (\rtan)  \op{P}_{lm} \pu \label{eq:locsplit}
\end{eqnarray}
\end{widetext}
This way, the ionic pseudopotential has been separated  into a local $l$-independent part and non-local operator. 
The local potential must decay according to \EQ{eq:potdecay}, but is otherwise arbitrary. A good choice is to take one of the $l$-channels of the atomic pseudopotential.  Since the local potential and all the $l$-components of the potential have the same long-range behavior, this formal transformation leads to $l$-dependent potentials that differ from zero only within spheres with radius $r_{\alpha l}^{\rm cut}$. This is numerically a decisive advantage. Atomic pseudopotentials are constructed up to a certain maximum angular momentum $l_{\rm max}$. From \EQ{eq:locsplit} it is clear that the truncation to $l_{\rm max}$ only applies to the non-local parts of the potential. One could therefore argue that higher angular momentum component are incorporated in the local potential. This fact is not effectively used in the construction of ionic pseudopotentials but will have some importance for the AEPs we will derive. 
 
For the implementation of the projectors we follow the separable form given by Kleinman and Bylander \cite{kleinman82a}:
\begin{equation}
\label{eq:Vkb}
\hat V^{KB}=V^{\rm loc}+\sum_{lm} \ket{\chi^{KB}_{lm}} E^{KB}_l\bra{\chi^{KB}_{lm}} \ko
\end{equation}
with the Kleinman-Bylander eigenvalue given by:
\begin{equation}
\label{eq:EKB}
E^{KB}_l=\frac{\langle\delta V_l u_l|\delta V_l u_l\rangle}{\langle u_l|\delta V_l|u_l\rangle} \ko
\end{equation}
and $\chi^{KB}_{lm}$ the normalized Kleinman-Bylander projectors
\begin{equation}
\label{eq:KB-projector}
\ket{\chi^{KB}_{lm}}=\frac{|\delta V_l \phi_{lm}\rangle}{(\langle u_l\delta V_l|\delta V_l u_l\rangle)^{1/2}} \pu
\end{equation}
The effective crystal potential from \EQ{eq:effpot} is now given more specifically as:
\begin{eqnarray}
\widehat{V}^{\rm eff} & = & V^{\rm psp,loc} + V^{\rm Hartree}[n] + V^{\rm xc}[n] +  \\
&& \sum_{lm} \ket{\chi^{KB}_{lm}} E^{KB}_l\bra{\chi^{KB}_{lm}}  \pu
\end{eqnarray}
During the self-consistent cycle of the Kohn-Sham equations (\EQ{eq:ks}) the density is updated until the ground state density $n^{\rm scf}$ is found. The starting point for the derivation of our AEPs is the local part of the self-consistent effective potential:
\begin{equation}
V^{\rm loc, eff}(\bm r) = V^{\rm psp,loc}(\bm r) + v_{\rm Hartree}[n^{\rm scf}] + v_{\rm xc}[n^{\rm scf}]  \pu
\end{equation}
with 
\begin{equation}
V^{\rm psp,loc}(\bm r) = \sum_{\alpha n} v^{\rm psp,loc}_{\alpha } (\rtan) \ko
\end{equation}
where $v^{\rm psp,loc}_{\alpha }$ is the local potential we defined in Eq.~\eqref{eq:vpsp} and 
 is an output of the pseudopotential construction procedure (e.g. after Hamann, Schl\"uter and Chiang \cite{hamann79a}, Kerker \cite{kerker80a}, Vanderbilt
\cite{vanderbilt85a} or Troullier and Martin\cite{troullier91a}) and usually includes one of the angular momentum component of the ionic pseudopotential. The Hartree and exchange-correlation parts are, however, non-spherically symmetric.

From the self-consistent real space potential we can move to reciprocal space using:
\begin{equation}
V^{\rm loc,eff} \argg = \frac{1}{\Omega_c}\int_{\Omega_c} V^{\rm loc,eff}({\bm r})\e{-i {\bm G} {\bm r}} \difr \pu
\end{equation}
We rewrite the local effective potential as a sum of atom-centered potentials:
\begin{equation}
V^{\rm loc,eff}({\bm r}) =  \sum_\alpha^{N_{\rm species}}  \sum_{n}^{N_\alpha} v_\alpha(\bm r - \basis)  \pu \label{vloc0}
\end{equation}
With this transformation we can rewrite the total reciprocal space potential (via variable substitution) as a Fourier sum:
\begin{eqnarray}
\label{eq:vlocstandard}
V^{\rm loc,eff} \argg &=&  \frac{1}{\Omega_c}\sum_\alpha^{N_{\rm species}}  \sum_{n}^{N_\alpha} 
 \e{-i {\bm G} \cdot {\bm \tau}_{\alpha n} } v_\alpha(\bm G)\quad 
 \end{eqnarray}
 with
 \begin{eqnarray}
v_\alpha(\bm G)&=& \int_{\infty} v_\alpha(\bm r)\e{-i {\bm G} {\bm r}}  \difr \pu
\end{eqnarray}

\section{AEP Construction Procedure}

\subsection{Binary systems}

The AEPs of a binary system cannot be directly extracted from the screened local effective potentials of DFT, but can be determined through $v_{+}$ and $v_{-}$, 
\begin{eqnarray}
v_{+} &=& v_{a} + v_{c} \ko \label{vpvm} \\
v_{-} &=& v_{a} - v_{c} \ko \nonumber
\end{eqnarray}
where $v_{a} (v_{c})$ stands for the anion (cation) potential. In the following we will use a binary system as prototype, but the procedure can be simplified to pure material by setting $v_{a} = v_{c}$. We start by explicitly dividing Eq.~(\ref{vloc0}) into cation and anion components:
\begin{equation}
 V^{(1)}_{\rm loc}({\bm r}) = \sum_{i=1}^{N_{a}}v_a({\bm r}-{\bm \tau}_i) + \sum_{j=1}^{N_{c}} v_c({\bm r}-{\bm \tau}_j) \ko \label{Vloc1} 
\end{equation}
with $N_a(N_c)$ as the number of anion (cations) in the structure.

Keeping the same atomic structure, and merely interchanging the atom types (index $i$ and $j$), we rewrite:
\begin{equation}
 V^{(2)}_{\rm loc}({\bm r}) = \sum_{i=1}^{N_{c}} v_c({\bm r}-{\bm \tau}_i) + \sum_{j=1}^{N_{a}} v_a({\bm r}-{\bm \tau}_j)  \pu \label{Vloc2}  
\end{equation}
By adding and subtracting $V^{(1)}_{\rm loc}$ and $V^{(2)}_{\rm loc}$, we find the following 
expressions for  $v_{+}$ and $v_{-}$:
\begin{eqnarray}
 V^{(1+2)}_{\rm loc}({\bm  r}) = V^{(1)}_{\rm loc}({\bm  r}) +  V^{(2)}_{\rm loc}({\bm  r}) &=& \sum_{n=1}^{N_{atoms}} v_{+}({\bm  r}-{\bm \tau}_n) \ko \\
 V^{(1-2)}_{\rm loc}({\bm  r}) = V^{(1)}_{\rm loc}({\bm  r}) -  V^{(2)}_{\rm loc}({\bm  r}) &=& \sum_{n=1}^{N_{atoms}} (-1)^{n+1}v_{-}({\bm  r}-{\bm \tau}_n) \pu
\end{eqnarray}
The Fourier transform of the last two equations leads to the relations:
\begin{eqnarray} 
 V^{(1+2)}_{\rm loc}({\bm  G}) &=& \frac{1}{\Omega}\int_{\Omega}e^{i{\bm  G}\cdot{\bm  r}}V^{(1+2)}_{\rm loc}({\bm  r})d{\bm  r} \nonumber \label{ftVp}\\
 &=& \frac{1}{\Omega}\left[\sum_{n=1}^{N_{atoms}} e^{i{\bm  G}\cdot{\bm \tau}_n}\right] v_{+}({\bm  G}) \ko \\
 V^{(1-2)}_{\rm loc}({\bm  G}) &=& \frac{1}{\Omega}\left[\sum_{n=1}^{N_{atoms}} (-1)^{n+1}e^{i{\bm  G}\cdot{\bm \tau}_n}\right]v_{-}({\bm  G}) \pu \label{ftVm} 
\end{eqnarray}
The potentials $v_{+}({\bm  G})$ and $v_{-}({\bm  G})$ are complex, but into the scheme of the spherical approximation used in this work, only the real part of the atomic potentials is of interest. We can extract the spherically averaged AEPs from Eqs. (\ref{ftVp}) and (\ref{ftVm}), by separating them into their real and imaginary components, and write:
\begin{eqnarray}
&& v_{\pm}^{\rm SA}(\left|G\right|) =  Re[v_{\pm}({\bm  G})] = \\ 
&& \Omega\left[\frac{Re[V^{(1\pm2)}_{\rm loc}({\bm  G})]}{\beta_{\pm}} + 
 \frac{Im[V^{(1\pm2)}_{\rm loc}({\bm  G})]}{\alpha_{\pm}}\right]\left(\frac{\beta_{\pm}\alpha_{\pm}}{\beta_{\pm}^2 + \alpha_{\pm}^2}\right) \ko \label{Vpm}
\end{eqnarray}
where
\begin{eqnarray} 
&&\beta_+ = \sum_{n=1}^{N_{atoms}}\sin({\bm  G}\cdot{\bm \tau}_n) \ko \\
&& \alpha_+ = \sum_{n=1}^{N_{atoms}}\cos({\bm  G}\cdot{\bm \tau}_n) \ko \nonumber \\
&&\beta_- = \sum_{n=1}^{N_{atoms}}(-1)^{n+1}\sin({\bm  G}\cdot{\bm \tau}_n) \ko \\
&&\alpha_- = \sum_{n=1}^{N_{atoms}}(-1)^{n+1}\cos({\bm  G}\cdot{\bm \tau}_n) \pu \label{alphas}
\end{eqnarray}

With $v_{\pm}^{\rm SA}(G)$, we can use Eqs. (\ref{vpvm}) to obtain the atomic potentials $v_{a}$ and $v_{c}$ in ${\bm G}$-space. 

The previous development can be used with any binary system. However, not every structure is suitable to extract the AEPs. It is, for example, of interest to obtain the potential for a high enough density of points, not only because that leads to the inclusion of the long range interaction, but also because that allows the utilization of a simple interpolation of points as a mean of constructing a continuous representation of the potentials in ${\bm G}$-space. 

To generate the atomic potentials, we used a zincblende structure made of 24 atoms and elongated along the [100] direction, regardless of the ground state structure of the different binary systems (wurtzite (wz), zincblende (zb) or rocksalt (rs)). We apply a compressive and a tensile deformation of 5\% along the slab direction in order to break the symmetry of the crystal and allow the extraction of the long range interaction. Note that a large supercell with ideal bulk atomic positions leads to a fine $\bm G$-vector mesh, but the structure factor is exactly zero at all the points, except at the $\bm G$-vectors of the bulk unit cell. As it should be, the large periodic supercell carries the same information as the unit cell, hence the need for the deformation. 

In Eqs. (\ref{ftVp}) and (\ref{ftVm}) we have assumed that the atomic potentials are independent of the atomic positions, which is strictly true for the bulk unit cell calculation but only approximately for the slab calculation. The main purpose to use the deformed slab calculation is to obtain points for small $\bm G$-vectors that are unavailable from a bulk unit-cell calculation. The slab calculations deliver, however, results for the entire range of $\bm G$-vectors and these results are identical (up to numerical accuracy) with the results obtained from a bulk unit-cell calculation for the large $\bm G$-vectors. For the shortest available $\bm G$-vector from a unit-cell calculation, $G_c$, we observe a small deviation between the slab and the uni-cell results. We therefore correct the slab results at this specific $\bm G$-vectors by a Gaussian correction to exactly obtain the unit-cell result (where the assumption of unique atomic potentials is exactly fulfilled).  
The shape of the Gaussian correction is fully defined by assuming that the correction has dropped to 1\% of its original value at $G$ equal to two times $G_c$. This procedure is not strictly necessary but slightly improves the results and requires only little effort.    

The orientation of the slab is chosen along the [100] direction because it allow for the extraction of $v_{-}^{\rm SA}(\left| \bm G\right|)$. This becomes clear if we write the local potential as 
\begin{equation}
 V_{\rm loc}({\bm r}) = \sum_n^{N_{atoms}/2} \{v_a({\bm r}-{\bm R}_n+{\bm r}_n) + v_c({\bm r}-{\bm R}_n - {\bm r}_n) \} \ko   
\end{equation}
with ${\bm r}_n$ the position of the atom $n$ measured from the anion-cation bond center ${\bm R}_n$. In reciprocal space we obtain, 
\begin{eqnarray}
V_{\rm loc}({\bm G}) &=& \frac{1}{\Omega}\sum_n^{N_{atoms}/2} e^{i{\bm G}\cdot{\bm R}_n} \{e^{-i{\bm G}\cdot{\bm r}_n}v_a(G) + e^{i{\bm G}\cdot{\bm r}_n} v_c(G) \} \label{VgRr}\\
 &=& \frac{1}{\Omega}\sum_n^{N_{atoms}/2} e^{i{\bm G}\cdot{\bm R}_n} \{\cos({\bm G}\cdot{\bm r}_n)v_+(G) \\
 && -i\sin({\bm G}\cdot{\bm r}_n)v_-(G) \} \pu \nonumber
\end{eqnarray}

We see, in the last equation, that whenever ${\bm G}$ and ${\bm r}$ are perpendicular (as can happen in the [110] orientation, where the cell can be written by placing the atoms in planes perpendicular to the direction of the structure) the quantity multiplying $v_-(G)$ becomes zero and $v_-(G)$ disappears from the equations. On the other hand, this expression shows us that $v_-(G)$ becomes less and less important as $G$ approaches zero.  

To summarize, the generation of AEPs comprises four steps: i) The DFT-LDA calculations of two 24 atoms distorted cell structures, where the atomic  positions are interchanged. ii) The extraction of the effective atomic potentials from the screened local potentials obtained in step (i), by applying Eq. (\ref{Vpm}) for $G$-vectors along the slab direction leading to a high density of $G$-points. iii) Adjustment to the bulk properties in such a way that $v_{\pm}^{\rm SA}(\left|G\right|)$ passes exactly trough $v_{\pm}^{\rm SA}(\left|G_c\right|)$. In this step a DFT calculation of the bulk system is needed. iv) A splines interpolation of the calculated points, in order to have a continuous form of the potentials in ${\bm G}$ space.\\

\subsection{Heterostructures and band-offsets}
\label{sec:linking}

Until now, the AEPs were derived from  DFT calculations of periodically repeated simulation cells of one material. For periodic systems, the average effective crystal potential, which corresponds to the $V(\bm G = \bm 0)$ component of the potential is ill defined and fixed to an arbitrary values in DFT codes. This means, that the energetic position of the eigenstates can be arbitrarily shifted along the energy axis. This also means that the eigenvalues of two DFT calculations for different materials, or even different structures of the same material, cannot be compared on an absolute scale. Only differences of eigenvalues within one DFT calculation can be formally compared (see, e.g,. Ref.\onlinecite{bester01b} for a discussion). In a heterostructure made of two materials A/B and treated with periodic boundary conditions, the $V(\bm G = \bm 0)$ component remains ill defined and is arbitrary. However, the difference between the ``local" average potential in the A and in the B regions, as far as this quantity can be meaningfully defined and calculated, represent the band-offset and is a very real and important quantity. The sign of the band-offset between materials will lead the the localization of the carriers in either the A-- or B--part of the structure. Calculations of band offsets based on DFT \cite{baldereschi88,wei98,shaltaf08,li09} involve the calculations of heterostructures and is a rather subtle task.

We incorporate the effects of band-offsets in our AEPs by linking them together using DFT calculations of heterostructures. The procedure involves the interchange of the cation anion positions within a QW heterostructure formed by two slabs  of different materials (A and B). The local potentials can be written as
\begin{eqnarray}
 V^{(1)}_{\rm loc}({\bm r}) &=& \sum_{i=1}^{N^A_{a}} v^A_a({\bm r}-{\bm \tau}_i)  + \sum_{j=1}^{N^A_{c}} v^A_c({\bm r}-{\bm \tau}_j)  + \nonumber \\ 
   & &\sum_{k=1}^{N^B_{a}} v^B_a({\bm r}-{\bm \tau}_k)  + \sum_{l=1}^{N^B_{c}} v^B_c({\bm r}-{\bm \tau}_l)  \ko \label{VlocAB1} \\ 
 V^{(2)}_{\rm loc}({\bm r}) &=& \sum_{i=1}^{N^A_{c}} v^A_c({\bm r}-{\bm \tau}_i)  + \sum_{j=1}^{N^A_{a}} v^A_a({\bm r}-{\bm \tau}_j)  + \nonumber \\
  & &\sum_{k=1}^{N^B_{c}} v^B_c({\bm r}-{\bm \tau}_k)  + \sum_{l=1}^{N^B_{a}} v^B_a({\bm r}-{\bm \tau}_l)  \pu  
\end{eqnarray}
Extracting the real part of the potential for slab $B$ ($v^B_{\pm}$) we obtain two equivalent expressions 
\begin{widetext}
\begin{eqnarray} 
Re[v^B_{\pm}({\bm G})] = \frac{\Omega}{\alpha^B_{\pm}}Re[v^{(1\pm2)}_{\rm loc}({\bm G})] - \frac{\alpha^A_{\pm}}{\alpha^B_{\pm}}Re[v^A_{\pm}({\bm G})] + \frac{\beta^B_{\pm}}{\alpha^B_{\pm}}Im[v^B_{\pm}({\bm G})] + \frac{\beta^A_{\pm}}{\alpha^B_{\pm}}Im[v^A_{\pm}({\bm G})] \ko \\
Re[v^B_{\pm}({\bm G})] = \frac{\Omega}{\beta^B_{\pm}}Im[v^{(1\pm2)}_{\rm loc}({\bm G})] - \frac{\beta^A_{\pm}}{\beta^B_{\pm}}Re[v^A_{\pm}({\bm G})] - \frac{\alpha^B_{\pm}}{\beta^B_{\pm}}Im[v^B_{\pm}({\bm G})] - \frac{\alpha^A_{\pm}}{\beta^B_{\pm}}Im[v^A_{\pm}({\bm G})] \ko
\end{eqnarray}
\end{widetext}
where $\beta_{\pm}$ and $\alpha_{\pm}$ are given by Eqs. (\ref{alphas}), but restricted to each one of the slabs $A$ or $B$.

We therefore have a set of two equations and four unknowns. Fortunately, the imaginary components, which are basically the result of the deformation or loss of sphericity of the potentials due to the environment, vanish in most of the points and the expressions simplify to
\begin{eqnarray} 
Re[V^B_{\pm}({\bm G})] = \frac{\Omega}{\alpha^B_{\pm}}Re[V^{(1\pm2)}_{\rm loc}({\bm G})] - \frac{\alpha^A_{\pm}}{\alpha^B_{\pm}}Re[V^A_{\pm}({\bm G})] \ko \\
Re[V^B_{\pm}({\bm G})] = \frac{\Omega}{\beta^B_{\pm}}Im[V^{(1\pm2)}_{\rm loc}({\bm G})] - \frac{\beta^A_{\pm}}{\beta^B_{\pm}}Re[V^A_{\pm}({\bm G})] \pu
\end{eqnarray}
We now fix the AEPs of slab A to the potential derived from the bulk calculation and extract from either one of the equations the potentials for slab B.

In this paper, we link the potentials by fixing each potential once it has been linked and subsequently using it as a basis to connect the new potential (linking B to A, C to B, D to C, ...,  A $\leftarrow $ B $\leftarrow $C $\leftarrow $ D 
$\leftarrow \cdot\cdot\cdot$). We defined the linking order by using QWs with the lowest possible lattice mismatch among the studied materials. We used a 48 atoms zincblende QW structure, along the [100] direction, made of two slabs (A and B) of equal sizes, with a lattice constant equal to the average of the lattice constants of the two materials, including the interfaces, and without any relaxation. 
 
\subsection{Deformation Potentials}

The $v_+(\bm G = \bm 0)$ component of the potential does only shift the eigenspectrum along the energy axis and has no meaning as it comes out of a DFT calculation. However, this value influences the conduction and valence band deformation potentials individually. Again, the gap deformation potential is (as it should) independent of this value. We decide to fix the  $v_+(\bm G = \bm 0)$ to reproduce the DFT values of the valence band deformation potential according to Li {\it et al.} \cite{li06,li06b} for most of the materials. For SiGe, we averaged the values of pure Si and pure Ge. For SiC, we reproduced the absolute deformation potential reported by Lambrecht {\it et al.} \cite{lambrecht91}. In the case of MgO, we used the value reported by Zhu {\it et al.} \cite{zhu08}
The values we used are given in Table \ref{tab:defpot}.

\begin{table}
\caption{Absolute volume deformation potential (in eV) for the valence band maximum ($a_{v}^{VBM}$).}
 \label{tab:defpot}
 \begin{tabular}{|l|c|l|c|l|c|}\hline
  III-V      &   $a_{v}^{VBM}$  &  II-VI  & $a_{v}^{VBM}$ & IV-IV  & $a_{v}^{VBM}$  \\ \hline 
  InSb       &   1.61           &  CdTe   &     0.89      & Ge     &     2.23     \\
  AlSb       &   2.65           &  CdSe   &     0.90      & SiGe   &     2.30     \\
  InAs       &   1.79           &  ZnTe   &     0.99      & Si     &     2.38     \\
  GaSb       &   1.91           &  ZnSe   &     1.23      & SiC    &     4.30     \\
  InP        &   1.83           &  MgO    &     1.66      & C      &     2.16     \\
  AlAs       &   2.93           &  ZnO    &     0.48      &        &              \\
  GaAs       &   2.24           &         &               &        &              \\
  AlP        &   2.60           &         &               &        &              \\
  GaP        &   1.98           &         &               &        &              \\ \hline 
 \end{tabular}
\end{table}

\section{Assessment of the potential quality}

\subsection{Used norm-conserving pseudopotentials}

We started by generating a set of norm-conserving pseudopotentials to be used in the DFT calculations. The pseudopotentials used to calculate the screened local potentials in DFT are directly related to the generated AEPs. The AEPs have to be used together with the non-local part of the norm-conserving pseudopotentials in any future calculation. As our main interest in extracting AEPs is the reduction in computational time when studying semiconductor nanostructures, we used soft pseudopotentials generated with the FHI code\cite{fuchs99}, by including only up to $p$ electrons when possible, and going up to $d$ electrons when strictly required. In the case of the group III-V and group IV semiconductors, we always used the default FHI values for the real space cut-off radius of the different angular momentum channels. In the case of the II-VI compounds, we increased the real space cut-off radius as it leads to faster convergence with the enegy cut-off at a very modest cost in accuracy (see later).

\subsection{Comparison with DFT for bulk: errors induced by the  spherical approximation}

The first test of our potentials is against the bulk results obtained by a self-consistent DFT calculation naturally performed without spherical approximation. We refer to these calculations as ``DFT" in the first column in Table \ref{tableEg} where we report the value of the band gaps. In the second column, ``AEP", we report the results using our AEPs. The third column gives the energy difference between both approaches and is the result of the errors stemming from the spherical approximation. The last column shows the DFT relaxed lattice constants ($a_o$) used to generate the atomic potentials. We have, in general, errors in the energy band gap of few tenths of meV. With the exception of carbon, with an error of 160 meV, which is still small compared to the LDA band gap of over 5 eV. In Table \ref{tableEg}, we find the first tests of the transferability of some AEPs. In the case of CdSe, MgO and ZnO, we have calculated the bang gap of stable structures different than zincblende. We find that the deviations due to the spherical approximation is in the same order of magnitude as the deviations we obtain by changing the crystal structure. Note that InAs and InSb have a vanishing band gap in LDA, so that we chose to give the gap at the L-point in Table \ref{tableEg}.   

\begin{table}
\caption{Comparison of the single-particle band gap gap at the $\Gamma$-point or at the ${\rm L}$-point when indicated by (*). The bulk crystals have mostly the zincblende structure and, in a few cases, the rocksalt (rs) or wurtzite (wz) structures. For the wurtzite structure, both lattice parameters, $c$ and $a$, are given.}
 \label{tableEg}
 \begin{tabular}{|l|c|c|c|c|}\hline
  Compound   &   DFT (eV)  &   AEP (eV)  & $\Delta$ (meV) & $a_o$ (a.u.)\\ \hline  \hline
  III-V:     &             &               &              &       \\ \hline   
  InSb (*)   &   1.3970    &    1.4691     &    72        &   12.053    \\
  AlSb       &   1.5864    &    1.6444     &    58        &   11.478    \\
  InAs (*)   &   1.6993    &    1.7570     &    58        &   11.442    \\
  GaSb       &   0.5611    &    0.6569     &    96        &   11.211    \\
  InP        &   0.1921    &    0.2491     &    57        &   11.186    \\
  AlAs       &   1.6493    &    1.7080     &    59        &   10.719    \\
  GaAs       &   0.4182    &    0.4887     &    70        &   10.596    \\
  AlP        &   2.5867    &    2.6419     &    55        &   10.429    \\
  GaP        &   1.3570    &    1.4253     &    68        &   10.344    \\ \hline \hline
  II-VI:     &             &               &              &       \\ \hline
  CdTe       &   0.2914    &    0.2631     &    -28       &   12.174    \\
  CdSe       &   0.2441    &    0.2330     &    -11       &   11.563    \\
  CdSe (wz)  &   0.2969    &    0.2891     &     -8       &  (a) 8.182    \\
             &             &               &              &  (c) 13.34    \\
  CdSe (rs)  &   0.6528    &    0.6624     &     10       &   10.711    \\
  ZnTe       &   0.8550    &    0.8094     &    -45       &   11.364    \\
  ZnSe       &   0.9960    &    0.9774     &    -19       &   10.707    \\
  MgO        &   2.8588    &    2.8702     &    -11       &   8.8920    \\
  MgO (rs)   &   4.0744    &    4.1021     &     28       &   8.1560    \\
  ZnO        &   0.6566    &    0.6037     &    -53       &   8.7620    \\
  ZnO (wz)   &   0.7423    &    0.7586     &     16       &  (a) 6.200    \\
             &             &               &              &  (c) 10.11    \\ \hline \hline
  IV-IV:     &             &               &              &       \\ \hline
  Ge         &   0.1646    &    0.2493     &     84       &   10.695    \\
  SiGe       &   1.5704    &    1.5121     &    -58       &   10.514    \\
  Si         &   2.5731    &    2.6601     &     87       &   10.356    \\
  SiC        &   5.3334    &    5.3965     &     63       &   8.3680    \\ 
  C          &   5.6474    &    5.8077     &    160       &   6.6870    \\ \hline   
 \end{tabular}
\end{table}

\subsection{External and Internal Strain Situations}

In the following test, we selected one III-V, one II-VI and one group IV material, namely AlP, ZnSe and SiC, and studied the band gaps under different strain conditions. We start by studying the bulk materials under the effects of uniform expansion and compression and show the results in \FG{unifDefEg}. We see a quasilinear dependence in all cases, going from large band gaps, when the sample is compressed, to smaller band gaps in the expanded situation. The AEP results follow well the self consistent DFT calculations. In \FG{unifDefDta} we plot the deviations between the AEP and the DFT results. This deviation becomes important when the effective potential starts to be significantly modified by self-consistent effects. Until around 5\% the approximation of rigid overlapping potentials remains accurate. The value of 5\% is very large for a hydrostatic deformation and will not be encountered in any experimental situation. 
\begin{figure}
 \includegraphics[width=\figurewidth]{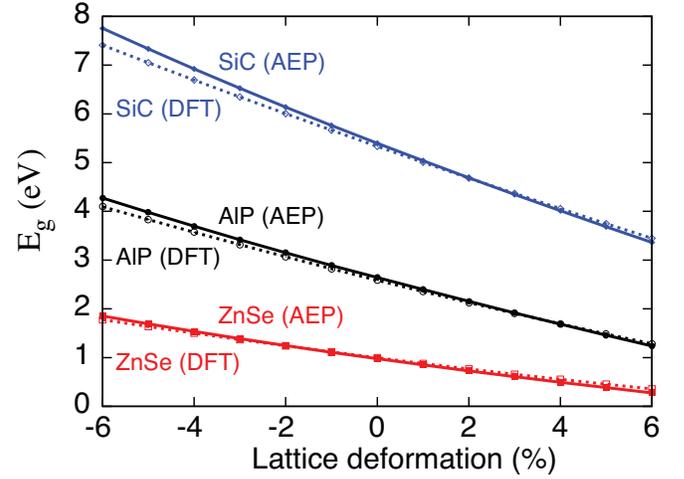}
\caption{Bandgap for selected bulk materials, as a function of the lattice constant deformation calculated using self-consistent DFT and using our AEPs. \label{unifDefEg}}
\end{figure}

Variations of the lattice constant are translated in terms of variations in the reciprocal space mesh. If the lattice constant is expanded, the $\bm G$-space mesh becomes denser and the AEP calculation probes $\bm G$-vectors that are shorter than the critical length $G_c$ defined by the relaxed lattice. The calculations done with compressed lattice constants accordingly probe $\bm G$-vectors longer than $G_c$. This constitutes a good indication of the transferability of our AEPs. 
\begin{figure}
 \includegraphics[width=\figurewidth]{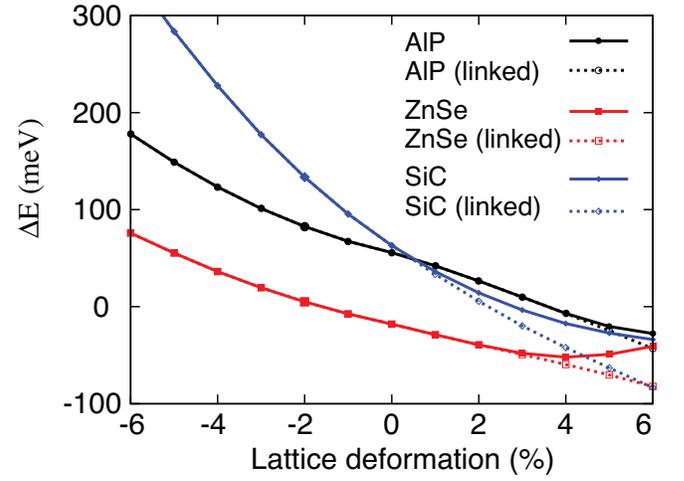}
\caption{Deviations of the AEP calculations from the DFT results for different bulk materials as a function of lattice deformation. \label{unifDefDta}}
\end{figure}

\FG{unifDefDta} also shows the deviations of the band gap energies when the calculations are done with the AEPs before and after (``link" notation in \FG{unifDefDta}) the linking procedure (Sec. \ref{sec:linking}). As expected, we see that the results of the AEP calculations are identical in the compressed situations. We have differences between the results of only few meV in the expanded condition, which is acceptable.

One more test performed in this section is aimed at testing not only the response of the material to strain, but also the long-range quality of our potential. For this purpose we constructed a 32 atom and a 36 atom cell elongated along the [110] and the [100] directions, respectively. We compress one half of the cell and expanded the other half, along their respective orientations. The lattice constant in the perpendicular direction was kept at the bulk value. With this construction, the interface between the expanded and the compressed regions has the ideal lattice constant as well, and the length of the cell remains constant for different deformations. 

\begin{table}
\caption{Energy band gap deviations of the AEP calculations from the DFT results $\Delta{\rm E}_g = {\rm E}_g{\rm (AEP)} - {\rm E}_g{\rm (DFT)}$. The AEPs are obtained from the binary system (Binary) and after the linking procedure (Linked). The deformation is applied along the [110] direction.}
 \label{table_110slb}
 \begin{tabular}{|c|c|c|c|c|c|c|}\hline
  Deformation &  \multicolumn{6}{|c|}{$\Delta{\rm E}_g$ (meV)}   \\ \cline{2-7}
              &  \multicolumn{2}{|c|}{AlP}  &  \multicolumn{2}{|c|}{ZnSe}     &  \multicolumn{2}{|c|}{SiC} \\ \cline{2-7}   
              &   Binary    &    Linked  &    Binary  &   Linked  &    Binary  &  Linked \\ \hline
  0\% (Bulk)  &   55.2      &   55.2    &    -18.6   &   -18.6   &     63.1   &    63.1   \\
  1\%         &   84.4      &   82.3    &    -20.7   &   -25.2   &     11.4   &   -34.1   \\
  2\%         &   81.6      &   76.1    &    -27.0   &   -39.1   &    -12.1   &   -118    \\
  3\%         &   81.4      &   72.0    &    -32.7   &   -52.1   &    -33.5   &   -208    \\
  4\%         &   83.4      &   69.5    &    -36.2   &   -62.1   &    -52.7   &   -304    \\
  5\%         &   87.8      &   68.7    &    -36.8   &   -68.7   &    -69.8   &   -405    \\
  6\%         &   94.6      &   69.7    &    -34.9   &   -72.2   &    -84.6   &   -511     \\ \hline 
 \end{tabular}
\end{table}

\begin{table}
\caption{Equivalent to Table \ref{table_110slb} for a deformation applied along the [100] direction.}
 \label{table_100slb}
 \begin{tabular}{|c|c|c|c|c|c|c|}\hline
  Deformation &  \multicolumn{6}{|c|}{$\Delta{\rm E}_g$ (meV)}   \\ \cline{2-7}
              &  \multicolumn{2}{|c|}{AlP}  &  \multicolumn{2}{|c|}{ZnSe}     &  \multicolumn{2}{|c|}{SiC} \\ \cline{2-7}   
              &   Binary    &    Linked      &    Binary  &   Linked  &    Binary  &  Linked \\ \hline
  0\% (Bulk)  &   55.2      &    55.2       &    -18.6   &   -18.6   &     63.1   &    63.1   \\
  1\%         &   81.7      &    72.5       &    -22.0   &   -26.5   &     8.30   &   -76.9   \\
  2\%         &   67.7      &    45.1       &    -32.6   &   -45.4   &    -36.1   &   -237    \\
  3\%         &   52.6      &    16.5       &    -44.1   &   -64.9   &    -82.4   &   -398    \\
  4\%         &   37.8      &   -11.4       &    -54.9   &   -82.6   &    -128    &   -558    \\
  5\%         &   23.6      &   -38.8       &    -63.7   &   -97.3   &    -173    &   -717    \\
  6\%         &   9.02      &   -66.5       &    -71.7   &   -110    &    -217    &   -793    \\ \hline 
 \end{tabular}
\end{table}

In Tables \ref{table_110slb} and \ref{table_100slb} we report the difference in the band gap between the AEP and the DFT calculations. As these slabs are binary systems, we can use again the potentials as originally derived from the deformed structures or the atomic potentials obtained after the linking procedure. In all cases the variations of the band gaps with the deformations are very well reproduced by the AEP calculations. However, with a slightly better agreement for the [110] slab. The deviations of the binary potentials of AlP and ZnSe (second and third columns in Table \ref{table_110slb}) change by only few meV as the deformation is increased. The better quality in the results of the [110] sample may be attributed to the fact that the potential $v_{-}$ does not play any role in this structure. In the case of SiC, the results are deteriorated by the linking procedure, which is due to the large lattice mismatch, of around 20\%, with respect to Si (Table \ref{tableEg}). We therefore have to use a strongly deformed structure to connect the potentials of SiC to Si and C, which compromises the quality of the results. In this case, the quality of the linked AEPs could be improved by selecting a more suitable structure to link the potentials. 

Since the deviations between the DFT and the AEP results are only slightly different for the two different orientations, we conclude that the long-range response of the potential to the strained situation is not significantly affected by the nature of the interface. In the case of a [110] structure, the interface layers are composed of mixed anion-cation atoms, while the interfaces in a [100] structure are composed of pure anion and pure cation atoms; qualitatively very different situations. This represents an important conclusion of our work.

\subsection{Semiconductor alloys: errors induced by the local average}

Here we considered 64 atom supercells of Ga$_x$Al$_{1-x}$As and Ga$_x$In$_{1-x}$P alloys in the zincblende structure. The alloying is simply done as a local average, where the AEP for an As atom surrounded by $n$ Ga and $(4-n)$ Al atoms, is written as a weighted sum of the type:
\begin{equation*}
\label{eq:sumofpot}
v_{\rm As} \{n {\rm Ga}, (4-n) {\rm Al}\} = \frac{1}{4} \left( n\; v_{\rm As}^{\rm GaAs} + (4-n)\; v_{\rm As}^{\rm AlAs} \right) \ko
\end{equation*}
where $v_{\rm As}^{\rm AlAs (GaAs)}$ represents the AEP of As derived from the AlAs (GaAs) structure. Note that the AEPs 
$v_{\rm As}^{\rm AlAs}$ and  $v_{\rm As}^{\rm GaAs}$ should be different since the As AEP contains informations on the bonding and it's environment. In practice both quantities are surprisingly similar.  
 In \FG{alloys} we show the DFT results along with the results of the AEPs in a similar way as done before. The atomic concentration has been modified in such a way that we have a successive transition from one bulk material to the other ($x=0$ or $x=1$). The atom positions have been relaxed in order to approach a realistic structure, by using the valence force field method \cite{keating66,williamson00}.

The AEP and DFT calculations show the same qualitative behavior and differ quantitatively by a magnitude similar throughout the composition range. These results demonstrate the validity of the linking procedure. It is worth mentioning here that AEP calculations using unlinked potentials lead to very different results as the band offsets are arbitrary.  
\begin{figure}
 \includegraphics[width=\figurewidth]{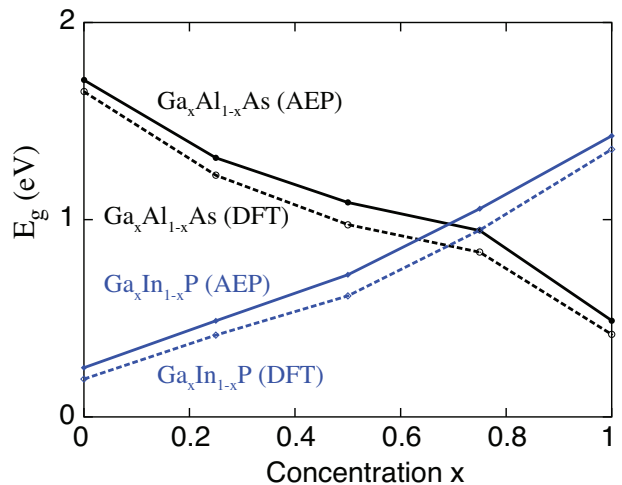}
\caption{DFT and AEP band gap calculations for a 64 atom alloy system at different concentrations. \label{alloys}}
\end{figure}

\subsection{Quantum wells: errors induced by a lack of charge transfer at the interface}

The next test we perform is aimed at testing the quality of the long-range response of the AEPs due to the presence of a heterostructure interface. In \FG{36atSLs} we plot the band gap as a function of the well (or barrier) thickness for a 36 atom [100] oriented A/B superlattice with zincblende structure. As test systems we select: GaP/AlAs, CdSe/ZnTe and Si/Ge. The structures have been relaxed by using and averaged lattice constant perpendicular to the superlattice direction and according to Poisson's ratio along the superlattice. 

Through this configuration procedure, the structures all have different dimensions, this means that they require a different set of $G$-vectors in reciprocal space. Besides, the $G$ mesh is rather dense, due to the size of the sample. This represents a highly demanding test for our AEPs since the full range of $G$-space is required. Furthermore, in each one of these calculations a considerable number of potentials is needed due to the presence of the interface. For instance, for the GaP/AlAs superlattice, we need eight AEPs; four for each one of the slabs (GaP and AlAs) and four for the two interfaces (AlP and GaAs).    

All the AEP calculations show good agreement with the DFT calculations in \FG{36atSLs}. The CdSe/ZnTe superlattice shows less deviations in the energy band gap, but the results are also less parallel to the DFT calculations in comparison to the GaP/AlAs and Si/Ge cases. This is due to the increase in the real space cut-off radii applied to the norm-conserving pseudopotentials of the II-VI materials, which leads to a loss in transferability in both, DFT and AEP calculations. However, the deviations for the CdSe/ZnTe superlattice never exceed 30 meV, well within our target accuracy.
\begin{figure}
 \includegraphics[width=\figurewidth]{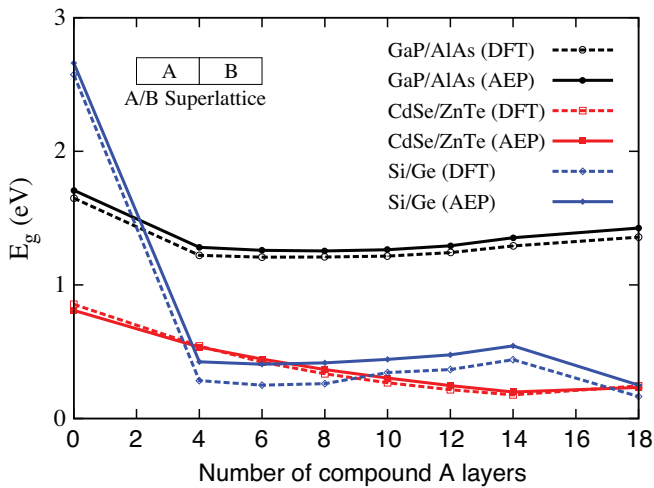}
\caption{DFT and AEP calculations for superlattices with 36 atoms along the [100] direction. The size of the wells and barriers are modified trough the variation of the proportion of the compound materials. \label{36atSLs}}
\end{figure}

In order to see the effects on the quality of the wave functions, we selected the GaP/AlAs supercell with 14 layers of GaP and 4 layers of AlP and show in \FG{wfk_GaPAlAsSL} the square of the VBM and CBM wave functions after averaging over the (100) planes. The figure shows a type II superlattice, where the VBM wave function is mostly localized in the GaP region and the CBM wave functions in the AlAs region. We compare the AEP with the DFT state densities and notice a good agreement for the envelope of these strongly oscillating functions, but also some differences in the values of maxima and minima within these atomic oscillations. Looking at the CBM densities carefully, we notice a slight increase of the density in the AlAs layer compared to the AEP case. A slightly less confined CBM wave function is in agreement with the small energetic differences found in \FG{36atSLs} for this case. Of relevance for the test is the fact that the long range response  of the potentials are called upon to describe the long-range oscillations of the envelope functions and the asymmetry on the localization of the wave functions due to the different character of the interfaces, GaAs on one side and AlP on the other.  
\begin{figure}
 \includegraphics[width=\figurewidth]{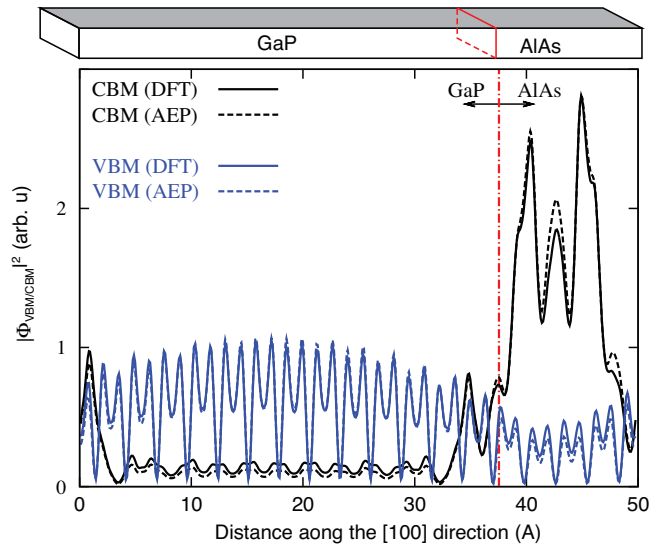}
\caption{Square of the CBM and VBM wave functions in a [100] GaP/AlAs superlattice with 14 layers of GaP and 4 layers of AlP. The wave functions have been integrated in planes along the axes of the superlattice. The calculations are done using DFT and our AEPs. \label{wfk_GaPAlAsSL}}
\end{figure}

\section{Summary}

In summary, we have developed a methodology to derived atomic effective pseudopotentials (AEPs) from simple DFT calculations. The procedure involves DFT calculations for two slabs with compressed and expanded regions. We establish an analytic connection between the DFT effective Kohn Sham potentials and the AEPs. The procedure is therefore free of parameters and does not involve any fitting procedure, which represents the main achievement of this work. We furthermore establish a method to connect the AEPs for different materials, rigorously from DFT calculations. Our AEPs intrinsically contain band-offsets between materials that are inherited from the DFT calculations, without having to explicitly calculate them. We demonstrate the accuracy and transferability of the AEPs for a total of 9 III-V, 6 II-VI and 5 group IV elements and find very good agreement with self-consistent DFT calculations. 

\begin{acknowledgements}
We would like to thank Roby Cherian for initial work on this project and Pierre-Yves Prodhomme for helpful discussions and 
acknowledge financial support by the German Science Foundation (DFG) Grant number 569323. 
\end{acknowledgements}

\end{document}